\documentclass[twocolumn,showkeys,9pt]{revtex4}
\usepackage{amsmath}
\usepackage{amsfonts}
\usepackage{amssymb}
\usepackage{epsfig}

\newtheorem{theorem}{Theorem}

\newtheorem{definition}[theorem]{Definition}

\newtheorem{lemma}[theorem]{Lemma}

\begin{document}

\title{Selfish Peering and Routing in the Internet}
\author{Jacomo Corbo$^1$}
\email{jacomo@eecs.harvard.edu}
\author{Thomas Petermann$^2$}
\address{$^1$ DEAS, Harvard University, 33 Oxford Street, Cambridge MA 02139, U.S.A. \\
$^2$ Laboratoire de Biophysique Statistique, ITP - FSB, \'{E}cole Polytechnique F\'{e}d\'{e}rale de Lausanne, CH-1015 Lausanne, Switzerland.}

\begin{abstract}
The Internet is a loose amalgamation of independent service providers acting
in their own self-interest. We examine the implications of this economic
reality on peering relationships. Specifically, we consider how the
incentives of the providers might determine where they choose to
interconnect with each other. We consider a game where two selfish network
providers must establish peering points between their respective network
graphs, given knowledge of traffic conditions and a nearest-exit routing
policy for out-going traffic, as well as costs based on congestion and
peering connectivity. We focus on the pairwise stability equilibrium concept
and use a stochastic procedure to solve for the stochastically pairwise
stable configurations. Stochastically stable networks are selected for their
robustness to deviations in strategy and are therefore posited as the more
likely networks to emerge in a dynamic setting. We note a paucity of
stochastically stable peering configurations under asymmetric conditions,
particularly to unequal interdomain traffic flow, with adverse effects on
system-wide efficiency. Under bilateral flow conditions, we find that as 
the cost associated with the establishment of peering links approaches zero, 
the variance in the number of peering links of stochastically pairwise stable equilibria increases dramatically.
\end{abstract}

\keywords{Internet Economics, Game Theory, Network Design, Network
Optimization}

\maketitle

\section{Introduction}

Much of the attention that has been paid to routing in data networks is
predicated on the assumption that the network is owned by a single operator.
In this scenario, the operator attempts to achieve some system-wide
performance objective like minimizing latency or minimizing
telecommunication costs. Such analyses still dominate, and yet a growing
number of network domains, like the Internet, consist of a loose federation
of autonomous, self-interested components, or network providers. In such a
world, the objectives of each individual provider remain the same but are no
longer necessarily consistent with any global performance measure. The
self-interested behavior of the parties involved means that the efficiency
of the whole network does not rely on an engineering solution per se, but is
inextricably tied to the economic realities of its implementation.

To understand the economic incentives endemic to the problem of
interconnecting networks, we must first characterize the nature of these
interconnections. Most relationships between two network providers can be
classified into one of two types: transit and peer \cite{norton}. Provider $A$
provides transit to provider $B$ if $B$ pays $A$ to carry traffic
originating within $B$ and destined elsewhere in the Internet (either inside
or outside $A$'s network). In such an agreement, provider $A$ accepts the
responsibility of carrying any traffic entering from $B$ across their
interconnection link.

In this paper, we are are primarly concerned with peering relationships.
Such interconnections consist in one or more bidirectional links established
between two providers $A$ and $B$. Unlike transit service, in a peering
relationship providers $A$ and $B$ will only accept traffic that is destined
for points within their respective domains, and there is no service level
agreement or monetary transfer between the two parties. This latter feature
means that peering decreases the reliance and therefore the cost of
purchased transit - which is the single greatest operating expense for Internet Service Providers (ISPs) \cite{norton}. Peering also lowers inter-Autonomous System (AS) traffic latency by reducing congestion at transit points, particularly National Access Points (NAPs) \cite{badasyan}.

But while peering has been a mainstay of Internet industry growth, for the
past several years, many ISPs have broken peering agreements because of
asymmetric traffic patterns and asymmetric benefits and costs from peering.
The reason for this stems from a 'tragedy of the commons' scenario that
arises when providers share a common backbone connection and pay no
penalties for overuse. A number of authors have made this point in a variety
of contexts \cite{norton,badasyan}. Such problems can be circumvented under transit
arrangements, however the prohibitive cost of monitoring Internet traffic
makes such agreements impractical - as illustrated by the relative paucity of
transit relationships between large backbones. Moreover, the benefits of peering, both among backbone (otherwise known as ``tier 1'') providers and between smaller ISPs, in reducing traffic latency telecommunication costs are well documented. For further details, see \cite{shin,badasyan}.

To understand the impact of peering relationships on network efficiency, we
now qualify their effects on network providers. When two providers form a
link connecting their networks (hereafter referred to as a \textit{peering
link}), the traffic flowing across that link incurs a cost on the network it
enters. Such a cost may be felt at the time of network provisioning (i.e.\ in
order to meet the quantity of traffic through a peering link, a provider may
have to increase its network capacity) or, alternatively, as an on-going network management
cost associated with coping with the increased congestion from additional
traffic. We avoid making any specific assumptions about the nature of the
network costs in our model, simply noting that these two interpretations
are possible.

One impetus for network providers' interconnection agreements is the value
gained by end users through that interconnection. Therefore, a complete
characterization of the economic incentives underlying interconnections must
include such benefits in addition to network costs. That said, in this paper
we work with a model introduced by Johari and Tsitsiklis \cite{johari} where two providers have already agreed to peer together. In doing so, we assume that the value to the end users is implicitly captured by this agreement. Our model only considers the network costs associated with peering relationships.

Consider a situation, then, where providers $A$ and $B$ are peers. Both
providers have a certain volume of traffic to send to each other and want to
minimize their costs. As mentioned earlier, because peering does not include
any service level agreement or monetary transfer, a tragedy of the commons
scenario emerges whereby each provider has a clear incentive to force
traffic into the other's network as quickly and cheaply as possible. This
phenomenon is known as ``nearest exit'' or ``hot potato'' routing and is the de
facto standard for outgoing traffic routing between peers. Nearest exit
routing's prevalence lies above all in the policy's simplicity - only local
knowledge is assumed - and its enforceability.

In this paper, we consider a problem that stems from the phenomenon of
nearest exit routing in interdomain peering. Given the distribution of
traffic between the two networks, both providers assume that the other will
use a nearest exit routing policy. The question then becomes the following:
Where (in their respective networks) will $A$ and $B$ like to establish
peering links? The decision of where to place their peering links is tied to
the providers' concern with minimizing network costs (whether provisioning
or congestion), which in turn is a function of the providers' network graphs
and the traffic distribution between them. Clearly, given the assumption of
nearest exit routing, the optimal placement for $A$ will not correspond to
that for $B$. We address the question of how
the differing preferences of the providers translate into a bilaterally negociated placement of
peering links. We are interested in understanding and
characterizing the networks that result when network providers choose their
peering connections in this way, as well as how the efficacy of these negotiated outcomes varies with
  cost and traffic flow parameters in the system.

Johari and Tsitsiklis \cite{johari} recently studied the peering point placement problem between two providers under the restriction of unilateral interdomain traffic flow, that is a special case of our model. They furthermore investigate the problem of optimally placing N peering links, and show 
that in the general case the optimal placement strategies for the sender and 
receiver providers are not the same. This result motivates our formulation of the problem as a game, where the number and location of peering links is 
endogenous to the model.

The paper is organized as follows. Section 2 formulates the peering point placement problem as a game theoretic model, also introducing the notions of pairwise stable and
  stochastically pairwise stable equilibria, as well as how the latter can be
  obtained via a stochastic process. In section 3, we describe the main findings regarding the artificial dynamics implied by our model and the distributions of the pairwise stable equilibria. We also mention preliminary results from ongoing work. Finally, conclusions are drawn in section 4.

\section{The peering point placement game}

In this section, we provide a description of the peering point placement
game we use to investigate the peer-connected networks that result when two
providers have already agreed to establish a peering relationship. The
network providers $A$ and $B$ consist of separate network graphs. We make
the assumption that $A$ and $B$ (both of size $N$ nodes) share the same
network topology. This strong assumption is justified on the grounds that
peering relations exist between similar-sized networks, e.g. between
backbone providers or between small ISPs; therefore we might expect some
similarity in their topologies. It is important to note that our results do
not hinge on this assumption: the paucity of peering equilibria we observe
in asymmmetric traffic conditions naturally lends itself to an
interpretation of asymmetrically sized networks. Mainly, we share this
assumption with a similar model presented in \cite{johari} in order to provide a point of comparison for our results, which is part of ongoing work.

We assume that both network providers send some amount of traffic to each
other and that both are using nearest exit routing. To provide a general
account of traffic distribution conditions, we fix the amount of traffic
sent from $A$ to $B$ to be $1$ packet from every node in $A$ to every node
in $B$, i.e. every node in $B$ receives $N$ packets from $A$. In the other
direction, we state that the amount of traffic sent from $B$ to $A$ is $%
\beta \in \lbrack 0,1]$ packets from every node in $B$ to every node in $A$.
The traffic distribution is therefore specified exogenously to the game
itself. However, note that as we increase $\beta $, we move away from the
completely unilateral situation where $A$ is the sole sender of traffic, to
the equally bilateral situation where $A$ and $B$ send each other an equal
volume of traffic.

Given these known traffic demands, network providers $A$ and $B$ play a game
to connect their two graphs so that traffic can be routed between them by
establishing some set of links $P\subseteq E$, where $E$ is the set of all
possible links between $A$ and $B$: $E=\left\{ (i,j):i\in A,j\in B\right\} $%
. $s_{ij}^{X}$ $\in \{0,1\}$ denotes the intention of a provider $X\in
\{A,B\}$ to establish a peering link between node $i\in X$ in its own graph
and node $j$ in the other's graph. Given the strategies of both providers,
given by the vector $s=(s_{ij}^{A},s_{ji}^{B},i\in A,j\in B)$, a
peer-connected network $g(s)=\left( A\cup B,\left\{ E_{A}\cup E_{B}\cup
P(s)\right\} \right) $ is formed, where
\begin{equation}
P(s)=\left\{ (i,j):s_{ij}^{A}\cap s_{ji}^{B},i\in A,j\in B\right\}
\end{equation}
where $E_{A}$ and $E_{B}$ denote the edges in the a priori defined graphs of 
$A$ and $B$, respectively. In words, a peering link is established between
nodes $i$ in $A$ and $j$ in $B$ if and only if it is desired by both
providers, i.e.\ if and only if $s_{ij}^{A}=s_{ji}^{B}=1$. The graph $g(s)$
then represents the entire peer-connected network. For notational
convenience, we sometimes refer to $g(s)$ simply as $g$ and $P(s)$ as $P$.

Given a set of peering links $P$, we assume that the routing of packets
results in a vector $\left( f_{i}^{X}(P),i\in X,X\in \{A,B\}\right) $, where 
$f_{i}^{X}(P)$ is the amount of flow passing through or terminating at node $%
i$ in graph $X$. This can be construed as the level of congestion at the
node $i$. We assume that $X$ incurs a cost $\alpha \in \lbrack 0,1]$ for
each unit of traffic which either passes or terminates at a node $i$, and a
cost $(1-\alpha )$ for every peering link in $P$. Therefore, given $g(s)$,
the total cost to network provider $X$ is
\begin{equation}
C_{X}(P(s))=\frac{\alpha }{n_{f}}\cdot \underset{i\in X}{\sum }%
f_{i}^{X}(P(s))+\frac{(1-\alpha )}{n_{p}}\left\vert P(s)\right\vert
\label{cost_function}
\end{equation}
where $n_{p}$ and $n_{f}$ are normalization factors: $n_{p}\geq 1$ is an
upperbound on the maximum number of links; $n_{f}$ is the worst-case
congestion for the network $A$ or $B$ \footnote{Recall that they are topologically identical.}. The calculation of $n_{f}$
hinges on our specification of $f_{i}^{X}$, which in turn depends on certain
flow conservation conditions. We avoid discussing these flow conservation
conditions here, instead referring the interested reader to a discussion of
monotonicity and flow feasibility conditions in \cite{johari2}. We simply point out that
under our flow assumptions, congestion is always diminished by the addition
of peering links and therefore maximized for some single peering link. In
our simulations, we conduct an exhaustive search to find this single
worst-case connection. We also let $n_{p}=N$, noting
that in equilibrium the number of outgoing links from a graph will never
exceed $N$. Finally, because we are modelling a situation where two
providers have already agreed to peer, we assure connectedness between $A$
and $B$ by imposing a very large penalty for disconnection.

\subsection{Pairwise stable equilibria}

For the following discussion, recall that $E$ is the set of all possible
links. For $P\subseteq E$, let $s_{P}$ denote the values of the strategy
vector $s$ restricted to the set of peering links $P$; that is,
\begin{equation}
s_{P}=\left( s_{ij}^{A},s_{ji}^{B},(i,j)\in P\right) .
\end{equation}
By an abuse of notation, we denote $s_{(i,j)}=s_{\{(i,j)%
\}}=(s_{ij}^{A},s_{ji}^{B})$. Therefore $s=\left( s_{(i,j)},s_{E\backslash
\{(i,j)\}}\right) $. The following definition describes the equilibrium concept to be studied thoughout the paper.

\begin{definition}
A strategy vector $s$ is pairwise stable if for every possible $(i,j)\in E$,
the following conditions hold:

(1) For any $s_{(i,j)}^{\prime }=\left( (s^{\prime
})_{ij}^{A},s_{ji}^{B}\right) :$%
\begin{equation}
C_{A}\left( s_{(i,j)}^{\prime },s_{E\backslash \{(i,j)\}}\right) \geq
C_{A}(s).
\end{equation}

(2) For any $s_{(i,j)}^{\prime }=\left( s_{ij}^{A},(s^{\prime
})_{ji}^{B}\right) :$

\begin{equation}
C_{B}\left( s_{(i,j)}^{\prime },s_{E\backslash \{(i,j)\}}\right) \geq
C_{B}(s).
\end{equation}

(3)For any $s_{(i,j)}^{\prime }=\left( (s^{\prime })_{ij}^{A},(s^{\prime
})_{ji}^{B}\right) $, at least one of the following holds:

\begin{equation}
C_{A}\left( s_{(i,j)}^{\prime },s_{E\backslash \{(i,j)\}}\right) \geq
C_{A}(s);
\end{equation}

\begin{equation}
C_{B}\left( s_{(i,j)}^{\prime },s_{E\backslash \{(i,j)\}}\right) \geq
C_{B}(s);
\end{equation}
\end{definition}
We will also refer to the network $g(s)$ generated by such a stategy vector $%
s$ as a pairwise stable network or a pairwise stable equilibrium.

The notion of pairwise stability, introduced by Jackson and Wolinsky \cite{jackson2}, is
meant to capture, in a static game setting, the dynamic process of
bargaining and negotiation which leads to the establishment of peering
links. Therefore, a link only remains in the graph if it is mutually
profitable for both link-constituting agents, while either party can decide
against any given link; i.e., link severance is unilateral while link
creation is bilateral. More formally, in checking whether a strategy vector $%
s$ is pairwise stable, Condition $3$ of Definition $1$ need not be checked
for $(i,j)\in P(s)$.
\begin{lemma}
Given a strategy vector $s$, suppose that:

(1) Conditions 1 and 2 of Definition 1 hold for $(i,j)\in E$; and

(2) If Condition 3 of Definition 1 holds for $(i,j)\notin P(s)$, then $s$ is pairwise stable.
\end{lemma}
Moreover, while pairwise stability is a weak stability notion, it is also
appealing because of its ability to generate sharp predictions about the
tension between stability and efficiency in many contexts \cite{jackson}. 

More generally, we might allow an agent to sever a subset of links $%
Q\subseteq P$, since this is a unilateral action. We make an important note
about pairwise stable equilibria in our game in this regard. We define a 
\textit{strong pairwise stable equilibrium} as a strategy vector $s$, or
equivalently a network $g(s)$, which is stable to the addition of single
links, as in Condition 3 of Definition 1, and the deletion of any subset $%
Q\subseteq P$ of links \cite{chakrabarti}. Ergo, the first two individual rationality
conditions of Defintion 1 have been strengthened.
\begin{lemma}
A strategy vector $s$ is pairwise stable if and only if it is strongly
pairwise stable.
\end{lemma}
This correspondence follows from the flow conservation (monotonicty and flow
feasibility) conditions in our model. Again, for a discussion of these
conditions, we refer the reader to \cite{johari2}. Therefore, while we keep referring to
pairwise stable equilibria, one should remember that such equilibria are
predicated on strong individually rational conditions in our game. Still
more generally, we might allow for simultaneous addition of links. This
would lead to a notion of stability accouting for coalitional deviations, which is beyond the scope of this work.

Johari and Tsitsiklis \cite{johari} show that finding the optimal placement of peering
links for either provider in a peering placement problem similar to the one
we have defined is NP-complete \footnote{They consider a situation where traffic flows unilaterally from one provider
to the other.}. In fact, solving for pairwise stable networks
suffers from the same problem of combinatorial explosion and is also
NP-complete (see \cite{garey} p.\ 206). NP-completeness suggests that all
known algorithms to solve the problem require time which is exponential in
the problem size \textit{(for instance, in the size of the network
providers' graphs)}. We cope with the intractability of our problem by
restricting our attention to the set of stochastically stable networks, i.e.
to the set of stochastically pairwise stable equilibria.

\subsection{Stochastically pairwise stable equilibria}

Given the intractability of providing a full characterization of pairwise
stable networks, we instead use a stochastic procedure to solve for a
distribution of stochastically pairwise stable networks. Our algorithm is
adapted from a dynamic process of network formation proposed by Jackson and
Watts \cite{jackson}.

Consider the following process: At each period, providers $A$ and $B$
consider either the addition or deletion of a single link, with equal
probability. If the providers consider the addition of a link, then some $%
(i,j)\in E$ is chosen at random and both providers independently decide
whether the addition of the link would be beneficial. The link is added if
it meets the approval of both players. This corresponds to Condition 3 of
Definition 1. If the providers consider the severance of a link, then some
link $(i,j)\in P$ is randomly chosen and both providers independently and
unilaterally decide whether the link in question should remain,
corresponding to Conditions 1 and 2 of Definition 1.

This procedure provides a mechanism whereby agents iteratively approach a
pairwise stable network, either terminating at such a network or in a fixed
cycle \cite{jackson}.

Now consider, a perturbed version of the above stochastic process where the
providers' correct decisions in creating, maintaining, and deleting links
are inverted with probability $\varepsilon \in \left( 0,a\right]$. These
incorrect appraisals may be understood as mistakes or mutations \cite{kandori}. The
characterization of the asymptotic behaviour of this process is due to Young
\cite{young} and Freidlin and Wentzell \cite{freidlin}.

Briefly, for small but non-zero values of $\varepsilon $, the perturbed
stochastic process denotes the traversal of an irreducible and aperiodic
Markov chain. Therefore, it has a unique limiting stationary distribution,
i.e.\ the process is ergodic. As $\varepsilon$ goes to zero, the stationary
distribution converges to a unique limiting stationary distribution. The
networks which are in support of that distribution are said to be \textit{%
stochastically stable}.

This process selects for networks with higher resistances, i.e. those with
larger basins of attraction. For $2\times 2$ games, the stochastically
stable states correspond to the risk-dominant equilibria \cite{harsanyi}.
Stochastically stable networks can therefore be construed as the pairwise
stable networks that are more likely to emerge in a dynamic process of
network formation. Furthermore, the above procedure provides an effective
way to characterize this set of networks via simulation.

\begin{figure}[b]
\includegraphics[width=9.5cm]{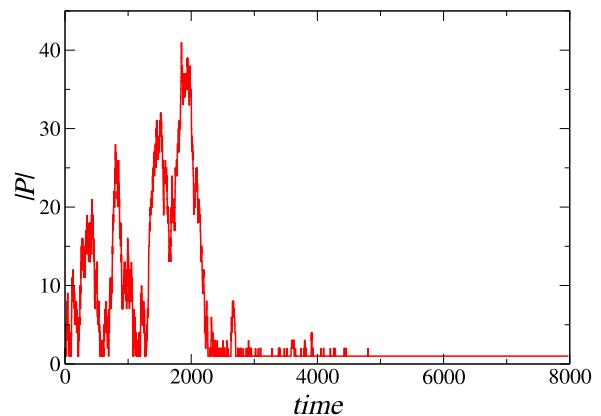}
\caption{Time evolution of the number of peering links for the unilateral case with cost parameter $\alpha=1$. After a certain equilibration period enforced by our stochastic algorithm, $|P|=1$ is reached.}
\label{fig:np_dyn}
\end{figure}

That said, we would like to make a cautionary note in this regard. The rate
of degeneration for $\varepsilon $ must be less than the slowest rate of
convergence to equilibrium for the unperturbed process \cite{kandori}, which makes large
state spaces difficult to search. The size of the state space in our game is
upperbounded by $2^{|E|}$ where $|E|=|A|^{2}$, given $A$ and $B$ are
topologically the same. Even so, simulations on graphs with as many as $100$
nodes indicate that the qualitative relations between parameters observed on
smaller graphs remain the same, suggesting the efficacy of the algorithm on
larger graphs (as well as the invariance of our results to topological
peculiarities).

We numerically simulate the unique limiting stationary distribution of the
perturbed dynamic process by the following simple rule:
\begin{equation}
\varepsilon ^{t}=\left\{ 
\begin{array}{c}
0.5\qquad \text{for} \qquad t<10000, \\ 
0.5\cdot e^{0.0001(10000-t)} \qquad \text{otherwise.}%
\end{array}%
\right. 
\end{equation}

\section{Results}

In this section we present some simulation results providing a partial
characterization of stochastically stable peering configurations when $A$
and $B$ are scale-free networks with $100$ nodes constructed according to the model based on growth and preferential attachment \cite{barabasi}. It is well known that the
topology of large ASes is closely scale-free.

\subsection{The unilateral case}

Fig.\ \ref{fig:np_dyn} shows the time evolution of the number of peering links for the case where $\alpha=1$ and $\beta=0$ (unilateral flow). The fact that quite many peering links are established for $t \lesssim 3000$ is due to the perturbed dynamic process. As far as the equilibrium state is concerned, we found $|P|_\text{eq}=1$ for $\alpha < 1$ and in the case $\alpha=1$, there was a $2\%$ chance of ending up with 2 final peering links.

\begin{figure}[h]
\includegraphics[width=9.5cm]{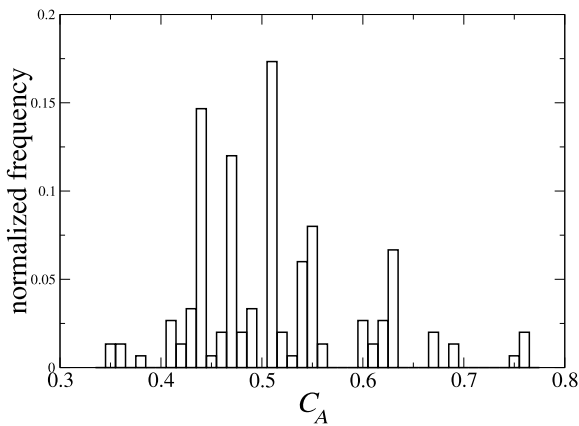}
\caption{Distribution of the sender's cost in the steady state (for unilateral flow). This result was obtained by statistically analyzing the costs of graph $A$ for times $10000 < t < 100000$ for 150 different runs.}
\label{fig:cs_unilat}
\end{figure}

When it comes to the costs, the receiver has to pay an amount of 0.3565 units independently of where the link is established. This has to do with the fixed topology and with the fact that the receiver ($B$) sends no packets in the present case. Concerning the sender, on the other hand, certain nodes are preferable to others in that the incurred traffic costs is lower. Fig.\ \ref{fig:cs_unilat} shows the sender's cost distribution in the stationary state.

\subsection{Bilateral flow}

\begin{figure}[h]
\includegraphics[width=9.5cm]{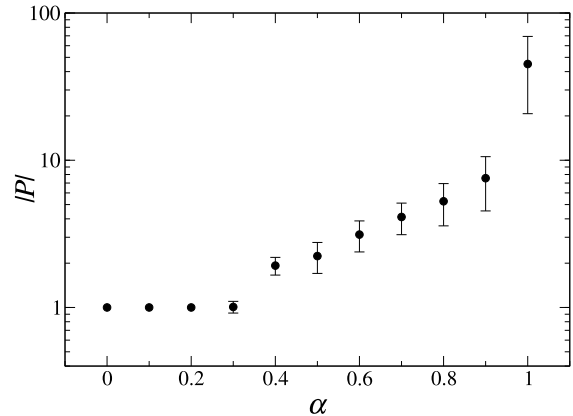}
\caption{The average number of established peering links as a function of $\alpha$ for the bilateral case. For $\alpha \leq 0.2$ only one link remained in equilibrium, hence the lack of the error bar. Up to $\alpha \leq 0.9$ the distributions were well peaked around the corresponding mean values whereas the present average values were scattered much more broadly for $\alpha=1$ (note the logarithmic scale). These results were obtained for times $10000 < t < 100000$ and for 150 different runs.}
\label{fig:np_ave}
\end{figure}
A more interesting situation arises when both ISPs $A$ and $B$ send data packets. We investigated the case where each node in graph $X$ sends one packet to every node in graph $Y$, $X,Y \in \{A,B\}$, i.e.\ $\beta=1$. Fig.\ \ref{fig:np_ave} shows the average number of peering links that were established in equilibrium. For $\alpha \leq 0.9$, these values are meaningful quantities as the accompanying variances appear to be small. Note also the exponential growth in the range $0.5 \leq \alpha \leq 0.9$. In the case where the peering links no longer contribute to the cost ($\alpha=1$), the peering structure is subject to much stronger fluctuations. In other words, there seems to be some type of phase transition for $\alpha \rightarrow 1^-$.

\begin{figure}[t]
\includegraphics[width=9cm]{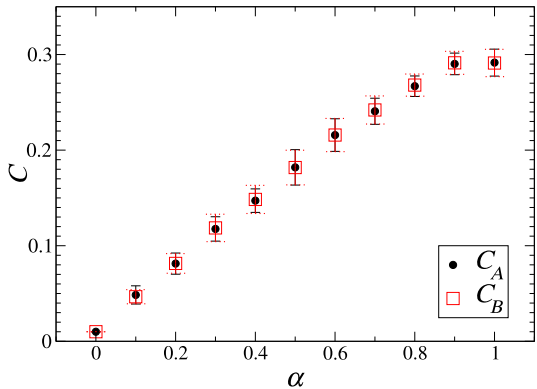}
\caption{$A$ and $B$'s costs (their averages) as a function of $\alpha$ for $\beta=1$. Note that the costs end up in well defined ranges, and the linear growth with $\alpha$ can be observed as well. For simulational details, see Fig.\ \ref{fig:np_ave}.}
\label{fig:cost_ave}
\end{figure}

Fig.\ \ref{fig:cost_ave} illustrates the mean values of the costs as $\alpha$ is varied. The fact that $C_A$ behaves very similarly to $C_B$ is a reflection of $\beta=1$. A rather linear relationship between $C$ and $\alpha$ can furthermore be observed for $\alpha \lesssim 0.8$.

\subsection{Ongoing work}

We are currently analyzing the equilibria of our game on smaller and more
regular, i.e.\ more tractable, topologies. Initial findings suggest that many
efficient peering configurations are pairwise stable over small ranges of $%
\alpha $ but only stable for a very precise $\beta $. Moreover, we note that
the set of pairwise stable configurations contain closely efficient graphs
for a broad range of $\alpha ,\beta $ pairs. Furthermore, for $\beta =1$ the
system exhibits a wide range of stochastically pairwise stable peering
configurations for intermediate values of $\alpha $. However, for $\beta <1$%
, there is a drastic paucity of stochastically stable equilibria and the
more efficient peering configurations are no longer stable. This suggests
that peering is sensitive to asymmetries, particularly in perceived traffic
load distributions. This also means that peering relations are more
sensitive to differences in traffic loads than to differences in network
size--the latter representing a variation in $\alpha $ as opposed to $\beta $%
. Of mention is that these results are irrespective of network topology. We
are currently working toward a full characterization of stable and efficient
configurations in our game.

\section{Conclusions}

We study how economic incentives affect the peering relationship between two
network providers. Specifically, we consider a game where two selfish
network providers must establish peering points between their respective
network graphs, given knowledge of traffic conditions and a nearest-exit
routing policy for out-going traffic, as well as costs based on congestion
and peering connectivity involving a parameter $\alpha$ which gives their relative importance. We focus on the pairwise stability equilibrium
concept and use a stochastic procedure to solve for the stochastically
pairwise stable configurations. 

We note a paucity of stochastically stable
peering configurations under asymmetric conditions, particularly to unequal
interdomain traffic flow, with adverse effects on system-wide efficiency. The volatility of
  peering relationships in the face of perceived asymetries suggests that
  peering will become increasingly rare as traffic and cost monitoring become
  more accurate and available.

For the case of equally bilateral traffic flow, we find a transition in behavior for $\alpha \rightarrow 1^-$, meaning that below this value, the number of peering links is well peaked around some mean value and above it, strong fluctuations are observed. We furthermore find that the costs of both providers grow linearly with $\alpha$.

\begin{acknowledgments}

We wish to thank to Kirk Doran for his valuable comments at the beginning of this work. One of us (T.P.) is also thankful to Mark Buchanan for suggesting this type of problem as well as to the FET Open Project IST-2001-33555 COSIN and to the OFES-Bern (CH) for partial financial support. We are furthermore very grateful to the Santa Fe Institute as well as to the Complex Systems Summer School 2004 faculty and participants for having provided a challenging and stimulating environment in which to produce this work.

\end{acknowledgments}


\begin{thebibliography}{99}

\bibitem{norton}
W.B. Norton. {\it Internet Service Providers and Peering.} Equinix White Papers (2001).

\bibitem{badasyan}
N. Badasyan and S. Chakrabarti. {\it Intra-backbone and Inter-backbone Peering Among Internet Service Providers.} e-print ewp-io/0407004 (2004).

\bibitem{shin}
S.-J. Shin and M. Weiss. {\it Internet Interconnection Economic Model and its Analysis: Peering and Settlement.} Netnomics {\bf 6}, 43 (2004).

\bibitem{johari}
R. Johari and J.N. Tsitsiklis. {\it Routing and Peering in a Competitive Internet.} Technical Report P-2570, MIT Laboratory for Information and Decision Systems (2003). 

\bibitem{johari2}
R. Johari, S. Mannor and J.N. Tsitsiklis. {\it A Contract-Based Model for Directed Network Formation.} Submitted to Games and Economic Behavior (2003).

\bibitem{jackson2}
M.O. Jackson and A. Wolinsky. {\it A Strategic Model of Social and Economic Networks.} J. Econ. Th. {\bf 71}, 44 (1996).

\bibitem{jackson}
M.O. Jackson and A. Watts. {\it The Evolution of Social and Economic Networks.} J. Econ. Th. {\bf 106}, 265 (2002).

\bibitem{chakrabarti}
S. Chakrabarti. {\it Middlemen in Peer-to-Peer Networks: Stability and Efficiency.} Proceedings of the 2nd Workshop on the Economics of Peer-to-Peer Systems, Harvard University, Boston MA (2004).

\bibitem{garey}
M.R. Garey and D.S. Johnson. {\it Computers and Intractability.} W.H. Freeman and Co., New York (1979).

\bibitem{kandori}
M. Kandori, G.J. Mailath and R. Rob. {\it Learning, Mutation, and Long Run Equilibria in Games.} Econometrica {\bf 61}, 29 (1993).

\bibitem{young}
H.P. Young. {\it The Evolution of Conventions.} Econometrica {\bf 61}, 57 (1993).

\bibitem{freidlin}
M.I. Freidlin and A.D. Wentzell. {\it Random Perturbations of Dynamical Systems.} Springer, New York (1998).

\bibitem{harsanyi}
J. Harsanyi and R. Selten. {\it A General Theory of Equilibrium in Games.} MIT Press, Cambridge MA (1988).

\bibitem{barabasi}
A.-L. Barab\'asi and R. Albert. {\it Emergence of Scaling in Random Networks.} Science {\bf 286}, 509 (1999).

\end{thebibliography}
\end{document}